\PassOptionsToPackage{doi=false,isbn=false,url=false}{bibtex}
\documentclass[letterpaper, 10 pt, conference]{ieeeconf} 

\IEEEoverridecommandlockouts                              

\usepackage{graphicx} 
\graphicspath{ {./figures/} }
\usepackage{amsmath} 
\usepackage{amssymb}
\usepackage{algorithm}
\usepackage[noend]{algpseudocode}
\usepackage{xcolor}
\definecolor{tumblue}{RGB}{0,101,189}

\usepackage{stfloats}
\usepackage{booktabs}
\usepackage{setspace}
\usepackage{tabularx}
\usepackage{booktabs}
\usepackage[T1]{fontenc}
\usepackage[scaled=0.85]{beramono}
\usepackage{listings}
\usepackage{caption}
\usepackage{cleveref}
\usepackage{url}

\Crefname{lstlisting}{Listing}{Listings}
\lstdefinestyle{sparqlstyle}{
  language=OCL,
  basicstyle=\footnotesize,
  stepnumber=1,
  numbersep=10pt,
  tabsize=2,
  showspaces=false,
  breaklines=true
}

\usepackage[nonumberlist,nopostdot,acronym,toc]{glossaries}
\newacronym{llm}{LLM}{Large Language Model}
\newacronym{mas}{MAS}{Multi-Agent System}
\newacronym{cps}{CPS}{Cyber-Physical System}
\newacronym{pa}{PA}{Product Agent}
\newacronym{ra}{RA}{Resource Agent}
\newacronym{api}{API}{Application Programming Interface}
\newacronym{m2m}{M2M}{Machine to Machine}
\newacronym{mqtt}{MQTT}{Message Queuing Telemetry Transport}
\newacronym{plc}{PLC}{Programmable Logic Controller}
\newacronym{admarms}{ADMARMS}{Axiomatic Design of a Multi-Agent Reconfigurable Mechatronic System}

\usepackage{balance} 

\makeatletter
\newcommand\fs@betterruled{%
  \def\@fs@cfont{\bfseries}\let\@fs@capt\floatc@ruled
  \def\@fs@pre{\vspace*{5pt}\hrule height.8pt depth0pt \kern2pt}%
  \def\@fs@post{\kern2pt\hrule\relax}%
  \def\@fs@mid{\kern2pt\hrule\kern2pt}%
  \let\@fs@iftopcapt\iftrue}
\floatstyle{betterruled}
\restylefloat{algorithm}
\makeatother

\author{Jonghan Lim$^{1}$, Birgit Vogel-Heuser$^{2}$, and Ilya Kovalenko$^{1}$
\thanks{$^{1}$Jonghan Lim and Ilya Kovalenko are with the Department of Industrial and Manufacturing and the Department of Mechanical Engineering, Pennsylvania State University, State College, USA
        (e-mail: \{jxl567; iqk5135\}@psu.edu).
        }%
\thanks{$^{2}$Birgit Vogel-Heuser is with the Institute of Automation and Information Systems,
        Technical University of Munich, Munich, Germany
        (e-mail: vogel-heuser@tum.de).
        }%
}

\title{\LARGE \bf
Large Language Model-Enabled Multi-Agent Manufacturing Systems
}

\begin{document}
\setlength{\textfloatsep}{2pt}

\maketitle
\thispagestyle{empty}
\pagestyle{empty}

\begin{abstract}

Traditional manufacturing faces challenges adapting to dynamic environments and quickly responding to manufacturing changes. The use of multi-agent systems has improved adaptability and coordination but requires further advancements in rapid human instruction comprehension, operational adaptability, and coordination through natural language integration. Large language models like GPT-3.5 and GPT-4 enhance multi-agent manufacturing systems by enabling agents to communicate in natural language and interpret human instructions for decision-making. This research introduces a novel framework where large language models enhance the capabilities of agents in manufacturing, making them more adaptable, and capable of processing context-specific instructions. A case study demonstrates the practical application of this framework, showing how agents can effectively communicate, understand tasks, and execute manufacturing processes, including precise G-code allocation among agents. The findings highlight the importance of continuous large language model integration into multi-agent manufacturing systems and the development of sophisticated agent communication protocols for a more flexible manufacturing system.

\end{abstract}

\section{Introduction}
\label{sec:introduction}

The escalating demand for customized products and the increasing complexities of production processes are driving a transformation in the manufacturing sector. This paradigm shift has required manufacturers to quickly adjust to evolving product specifications and adeptly address unforeseen operational changes. In this context, manufacturers are required to not only maintain high-quality products and cost efficiency but also be agile and responsive to a dynamically changing environment.

One significant approach for enhancing adaptability and quick response of manufacturing systems is using multi-agent control~\cite{vogel2015agents}. A \gls{mas} includes several autonomous agents for decision-making, communication, and coordination~\cite{wooldridge2009introduction}. Many multi-agent architectures have been developed for manufacturing systems~\cite{kovalenko2019model, bi2023dynamic}. By enabling decentralized decision-making and enhancing coordination among various agents, \gls{mas} have contributed significantly to handling complexities in the manufacturing system~\cite{kovalenko2022cooperative}. 

However, there are still challenges in \gls{mas} frameworks in manufacturing. While agents in \gls{mas} can automate processes and make decisions, their ability to adapt to new formats or specifications is limited, leading to increased downtime. Thus, a challenge includes the limited ability of agents to dynamically adapt to new information including product specifications \textit{(RCh1)}. Another challenge lies in processing unstructured textual data. Textual data is crucial because it is more expressive than numerical data, offering vast knowledge that is currently underutilized in manufacturing, where structured data fails to include varied influencing factors~\cite{may2022applying}. Natural language processing capabilities to formalize unstructured textual data can reduce downtime and optimize processes~\cite{may2022applying}. Therefore, natural language processing is essential for interpreting and implementing textual data from human operators or interconnected systems \textit{(RCh2)}. 

The advent of \glspl{llm} presents a novel solution to these challenges. \glspl{llm}, such as OpenAI's GPT-3~\cite{floridi2020gpt} and GPT-4~\cite{achiam2023gpt}, provides opportunities to bring more flexibility and adaptability. With \gls{m2m} protocols, such as \gls{mqtt}~\cite{hunkeler2008mqtt} and MTConnect~\cite{MTConnectInstitution} that uses predefined structures, incorporating \glspl{llm} offer benefits. \glspl{llm} can interpret, generate, and act on natural language instructions, facilitating complex decision-making processes and adaptive responses to changing manufacturing conditions~\cite{makatura2023can}. 
The integration of natural language processing capabilities may also benefit operators and other system users who have limited experience with the specific process or system~\cite{chen2024learning}. Natural language processing capabilities can allow new users to quickly adapt to managing and optimizing processes and resources~\cite{schulte2021expert}.
The main contributions of this paper include:
Leveraging \gls{llm} to allocate and execute manufacturing tasks based on user instructions to enhance task efficiency \textit{(Cn1)},
enabling \gls{llm} agents to adapt to evolving specifications ensuring the production process is executed with the appropriate resources \textit{(Cn2)},
and applying natural language processing capabilities to enable agents to comprehend context-specific instructions from users and other agents \textit{(Cn3)}.

The rest of the manuscript is organized as follows.
\Cref{sec:background} provides background regarding \gls{mas}, \glspl{llm} in \gls{mas}, and \glspl{llm} in manufacturing.
\Cref{sec:framework} details the functioning of the proposed framework.
\Cref{sec:casestudy} demonstrates a case study to illustrate the applications and effectiveness of our framework.
\Cref{sec:challenges} presents challenges and limitations while integrating \gls{llm}.
\Cref{sec:conclusion} summarizes the paper and discusses future work.

\section{Background}
\label{sec:background}

\subsection{Multi-Agent Systems in Manufacturing}
\label{subsec:llm-mas}

Various \gls{mas} architectures have been developed to enhance agent communication, significantly contributing to the increased manufacturing system flexibility. For instance, the \gls{admarms} framework facilitates design-based reconfigurability~\cite{farid2015axiomatic}. Studies by Wang et al.~\cite{wang2016towards} and Kim et al.~\cite{kim2020multi} integrated \gls{mas} with big data analytics and reinforcement learning. Additionally, Kovalenko et al.~\cite{kovalenko2019dynamic} and Bi et al.~\cite{bi2023dynamic} focused on dynamic resource and task allocation within \gls{mas}, with frameworks for resource task negotiation and a model-based resource agent architecture to address manufacturing disruptions.

While these \gls{mas} architectures were developed to enhance agent communication for increased flexibility, they focus on utilizing predefined algorithms for task allocation and agent communication. Furthermore, there has been limited exploration in using natural language interpretation from human operators or other agents to enhance the responsiveness of manufacturing process control. By leveraging \glspl{llm}, agents can improve operational flexibility and efficiency in addressing these gaps.

\subsection{Large Language Models in Multi-Agent Systems}
\label{subsec:llm-mas}

The emergence of \glspl{llm} has sparked interest in developing interacting autonomous agents powered by the \gls{llm} technology~\cite{wang2023survey}. \gls{llm}-enabled agents demonstrate enhanced problem-solving capabilities through multi-agent discussions and diverse applications. For instance, Park et al.~\cite{park2023generative} showcase a simulation with 25 generative agents in a virtual town, facilitating social understanding studies.
Another work by Wu et al.~\cite{wu2023autogen} introduces an open-source framework called \textit{AutoGen} that facilitates customizable agent interactions and conversation programming. ChatEval~\cite{chan2023chateval} uses debater agents in an \gls{llm}-powered \gls{mas} for text evaluation, achieving enhanced accuracy and depth over traditional methods and single agent setups. MetaGPT~\cite{hong2023metagpt}, a meta-programming framework, enhances multi-agent collaboration using an assembly line approach for software engineering tasks.

These works demonstrate advanced multi-agent interactions, realistic behavior simulation, and task efficiency, which can enhance adaptability, coordination, and advanced natural language processing capabilities in manufacturing. 
Building on these insights, we explore integrating \glspl{llm} in a multi-agent framework to address the unique requirements and benefits in the manufacturing sector.

\subsection{Large Language Models in Manufacturing}
\label{subsec:llm-mas}

With the advancement of \gls{llm} technology, recent studies have been increasingly exploring both the opportunities and challenges of implementing \glspl{llm} within the manufacturing domain. 
Makatura et al.~\cite{makatura2023can} explores the application of \glspl{llm} in design and manufacturing, including CAD/CAM. Their approach investigates the potential role of \glspl{llm} in a variety of tasks, including design specification and manufacturing instruction, and highlights the capabilities and limitations. The analysis of GPT-4 shows strengths in design and manufacturing knowledge but faces challenges in reasoning and accuracy, with potential solutions including specialized languages and better context management~\cite{makatura2023can}. Jignasu et al.~\cite{jignasu2023towards} provide a comprehensive evaluation of \glspl{llm} for understanding, debugging, and manipulating G-code in 3D printing, testing their capabilities in error detection, correction, and geometric transformations. Similar work by Badini et al.~\cite{badini2023assessing} explores using ChatGPT to optimize the G-code generation in additive manufacturing. These studies prove that \gls{llm} can interpret G-code, essential for tasks like debugging, manipulating, and comprehending the code. Fakih et al.\cite{fakih2024llm4plc} and Koziolek et al.~\cite{koziolek2023chatgpt} focus on enhancing programming in industrial control systems, specifically \glspl{plc}, with frameworks like LLM4PLC and studies assessing GPT-4's capability to generate control logic from natural language prompts.

Several studies have integrated \glspl{llm} into manufacturing, including CAD/CAM, G-code, and \glspl{plc}, to enhance efficiency. However, these studies have not focused on using \glspl{llm} with \gls{mas} for autonomous task allocation, adaptability to evolving specifications, and interpretation of context-specific instructions. The framework described in~\Cref{sec:framework} leverages \gls{llm} to enhance multi-agent interactions and decision-making for manufacturing processes. This framework develops autonomous agents that use context-specific natural language instructions to dynamically adjust to new product information and coordinate tasks in real-time. Our approach is an initial step toward enhancing the CAD/CAM workflow, automatically adapting CAM processes based on changes in CAD specifications, ensuring seamless manufacturing operations.

\section{Framework}
\label{sec:framework}

\begin{figure*}[t]
\smallskip
\smallskip
    \centering
    \captionsetup{belowskip=-15pt}
    \includegraphics[width=.93\textwidth]{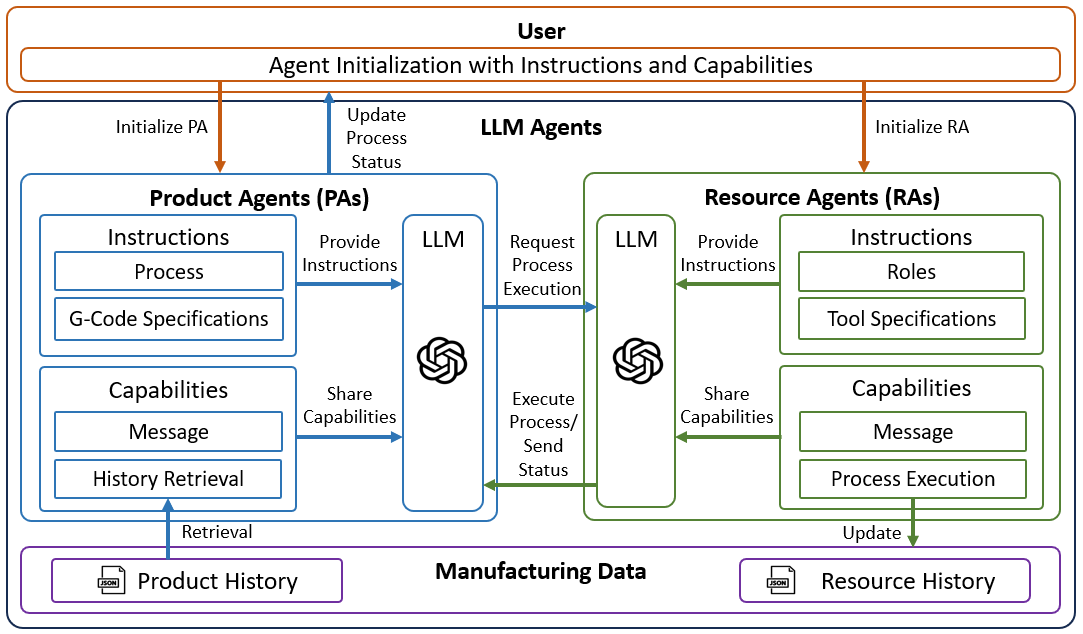}
    \caption{Framework for \gls{llm}-Enabled Multi-Agent Manufacturing Systems}
    \label{fig:architecture}
\end{figure*}

The novel architecture in \Cref{fig:architecture} is designed based on a \gls{mas} that includes both \glspl{pa} and \glspl{ra}. A \gls{pa} is an agent responsible for decision-making related to a product, while an \gls{ra} serves as a controller for a specific resource, such as a robot or machine~\cite{vrba2010rockwell}. 

\Cref{fig:architecture} provides an overview of an \gls{llm}-enabled multi-agent manufacturing system framework, where users initialize agents with instructions and capabilities. \glspl{pa} are assigned specific processes including G-code specifications, while possessing the ability to communicate and access historical data for informed decision-making. \glspl{ra} are responsible for the process execution of tasks, informed by instructions and tool specifications. \gls{llm} is applied for processing natural language inputs from both \glspl{pa} and \glspl{ra}. For evaluation, the chosen \gls{llm} is GPT-4, the latest and most advanced version of GPT developed by OpenAI~\cite{achiam2023gpt}.

This framework facilitates adaptability, enables real-time coordination, and provides the ability to respond to the demands of a complex manufacturing environment with natural language processing capabilities.

\subsection{Agent Initialization}
\label{subsec:initialization}

\begin{algorithm}
\caption{Agent Creation}\label{alg:agent_creation}

\begin{algorithmic}[1]
\algrenewcommand\algorithmicindent{0.7em}
\footnotesize{
\Procedure{CreateAgents}{$Agent\_list$, $Agent\_functions$}
    \State $Agents \gets []$
    \For{$agent\_initialization$ in $Agent\_list$}
        \State $configuration \gets$ LoadJSON($agent\_initialization$)
        \For{$name$, $agent\_data$ in $configuration$}
            \State $function\_list \gets []$
            \If{'functions' in $agent\_data$}
                \For{$function\_name$ in $agent\_data$['functions']}
                    \State $function\_reference \gets$ GetFunction($Agent\_functions$, 
                    \State \hspace{\algorithmicindent}$function\_name$)

                    \If{$function\_reference \neq$ None}
                        \State $function\_list$.append($function\_reference$)
                    \EndIf
                \EndFor
            \EndIf
            \State $agent \gets$ NewAgent($name$, $agent\_data$, $function\_list$)
            \State $Agents$.append($agent$)
        \EndFor
    \EndFor
    \State \Return $Agents$
\EndProcedure
}
\end{algorithmic}
\end{algorithm}

The agent initialization process configures a \gls{mas} by defining roles, capabilities, and communication protocols for each agent. This setup ensures agents operate effectively and collaborate towards system goals. The following initial prompt is given to all agents in the system:

\begin{quote}
{\footnotesize
\texttt{"You are a helpful agent in a cooperative Multi-Agent System.
If you are asked for a service you can provide you should help.
If necessary, you may ask the other agent for clarifying information.
You may communicate with your peers to achieve your goals.
If you do not know the answer do not make things up.
Only use the functions you have been provided with.
However, you may call these functions recursively."}
}
\end{quote}

These instructions ensure each agent operates within its capabilities, engages in effective communication, and contributes to the \gls{mas}'s objectives.

Following the initial prompts, the initialization process assigns specific roles to agents using various parameters. These parameters are selected to achieve adaptability, real-time response, autonomous decision-making, and efficient coordination within the \gls{mas}. Key parameters include:

\begin{itemize}
  \item \textbf{Functions}: Capabilities within the manufacturing that range from retrieving history to executing tasks.
  \item \textbf{Annotation}: Agent’s role and responsibilities, which aid in selecting the appropriate \gls{llm} agent.
  \item \textbf{Instructions}: User instructions offer a sequence of operations, techniques, and standards for agent activities.
  \item \textbf{Inbox}: The agent receives and manages instructions, updates, and feedback from other agents.
\end{itemize}
      
For \glspl{pa}, an additional parameter is introduced:

\begin{itemize}
  \item \textbf{Specification File}: Detailed manufacturing specifications, such as G-code files for CNC operations, for precise manufacturing execution.
\end{itemize}

\glspl{ra} are further configured with:

\begin{itemize}
  \item \textbf{Configurations}: This parameter outlines the operational settings specific to \glspl{ra}, such as process times, defect rates, and breakdown probabilities.
\end{itemize}

Integrating parameters into \gls{llm}-enabled \gls{mas} for manufacturing extends standardized approaches (e.g. FIPA~\cite{o1998fipa} and JADE~\cite{bellifemine1999jade}) by incorporating natural language processing for more flexible agent interactions. 
Integrating user-defined instructions and parameters in agent initialization can be further expanded through user studies to ensure industry applicability.

\begin{algorithm}
\caption{Chat Interaction}\label{alg:chat_interaction}
\begin{algorithmic}[1]
\algrenewcommand\algorithmicindent{0.7em}
\footnotesize{
\Procedure{chat}{$prompt$, $model$}
    \State $with\_functions \gets$ len(function\_info) $> 0$
    \State $msgs \gets []$
    \If{instructions}
        \State msgs.append(\{"role": "system", "content": instructions\})
    \EndIf
    \State msgs.append(\{"role": "user", "content": $prompt$\})
    \State $response \gets$ \_\_get\_response($msgs$, $model\_$, $with\_functions$)
    \State msgs.append($response$.message)
    \State $function \gets response$.message.function\_call.name
    \State $args \gets$ json.loads($response$.message.function\_call.arguments)
    \State $function\_response \gets$ executables[$function$](**$args$)
    \State \Return $function\_response$, $msgs$
\EndProcedure
}
\end{algorithmic}
\end{algorithm}

Based on the initialization information, the \texttt{Agent Creation} function, cp.~\Cref{alg:agent_creation}, dynamically generates agents. The \texttt{Agent Creation} function provides advanced \gls{llm} agents with different capabilities in manufacturing systems for adaptability to changing requirements. The algorithm iterates over each entry in the \textit{Agent\_list}, which contains initialization information. For every entry, configuration data \textit{configuration} for the agent is loaded from a JSON, containing details such as the agent's name, role, capabilities, and any specific functions. For each agent, an empty list \textit{function\_list} is prepared. If the \textit{`functions'} key is present, the algorithm iterates through each function name listed. For each function name, a reference \textit{function\_reference} is obtained by searching within a provided set of agent functions \textit{Agent\_functions}. If the function is found, its reference is appended to \textit{function\_list}. A new agent instance \textit{agent} is created with the specified name, agent data, and the list of function references. This agent is then added to the \textit{Agents} list. This algorithm facilitates the flexible creation of agents allowing for adaptation to various tasks.

\subsection{Agent Capabilities and \gls{llm} Integration}
\label{subsec:capabilities}

Through the integration of \gls{llm}, agents gain the capability to interpret complex instructions and specifications, offering a flexible approach to manufacturing operations. The framework 
uses \gls{llm} to decode user instructions for manufacturing and leverages the OpenAI \gls{api} feature called \textit{function calling}~\cite{function2024calling}. This feature allows for the automatic execution of user-defined functions, including specific manufacturing operations (e.g., milling, drilling) through descriptions that include operation parameters and required inputs like process name and product specifications (e.g. G-code).

The \texttt{`chat'} function, cp.~\Cref{alg:chat_interaction}, facilitates interaction among agents, based on available function information and executing function calls as defined by the manufacturing process requirements. The outcome, \textit{function\_response}, includes the details like operation type, product, and machine names, along with a message, \textit{msgs}, to facilitate the subsequent steps in the process. This approach enables context-aware execution of tasks, enhancing adaptability and efficiency. 

\subsection{Operational Workflow}
\label{subsec:workflow}

\begin{figure}[t]
\smallskip
\smallskip
    \includegraphics[width=.48\textwidth]{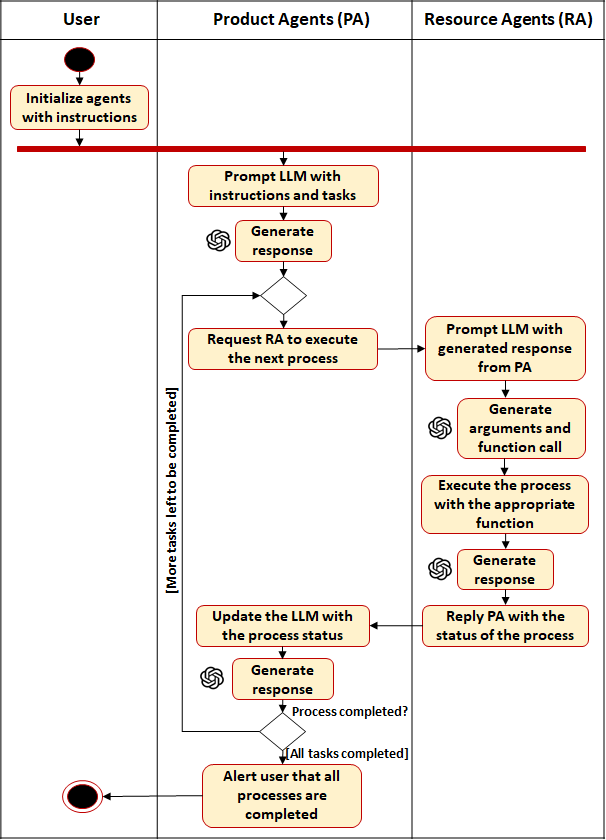}
    \caption{Operational Workflow of the Framework Using \gls{llm} for Task Execution and Coordination}
    \label{fig:activityDiagram}
\end{figure}

The workflow, as depicted in \Cref{fig:activityDiagram}, shows how user instructions in textual data can be converted into actionable operations. This approach enables the translation of high-level directives of users into machine-executable actions, enhancing the responsiveness to changing conditions.

The workflow begins with the user initializing \glspl{pa} and \glspl{ra}. Users describe product goals, sequences of operations, and specifications. This information is then processed by the \gls{llm}, which generates responses to guide the process. The \gls{pa} determines the suitable \gls{ra} for the next task, prompting it to carry out the required operation. The selected \gls{ra} constructs the necessary arguments to request the corresponding function call, leading to the execution of the specified process. Upon successful operation, the \gls{llm} formulates an appropriate response, informing the \gls{pa} of the process status. This feedback updates the \gls{llm}-integrated \gls{pa}  with the current status, determining the completion of the task. If additional tasks remain, the cycle repeats; otherwise, the user is notified of the completion of all processes.

\section{Case Study}
\label{sec:casestudy}

In this case study, we examine the proposed framework and demonstrate how the \gls{llm} is integrated into the framework to execute the manufacturing process.

\subsection{Case Study Setup}
\label{subsec:case-study-setup}

\begin{figure}[t]
\smallskip
\smallskip
    \centering
    \includegraphics[width=.48\textwidth]{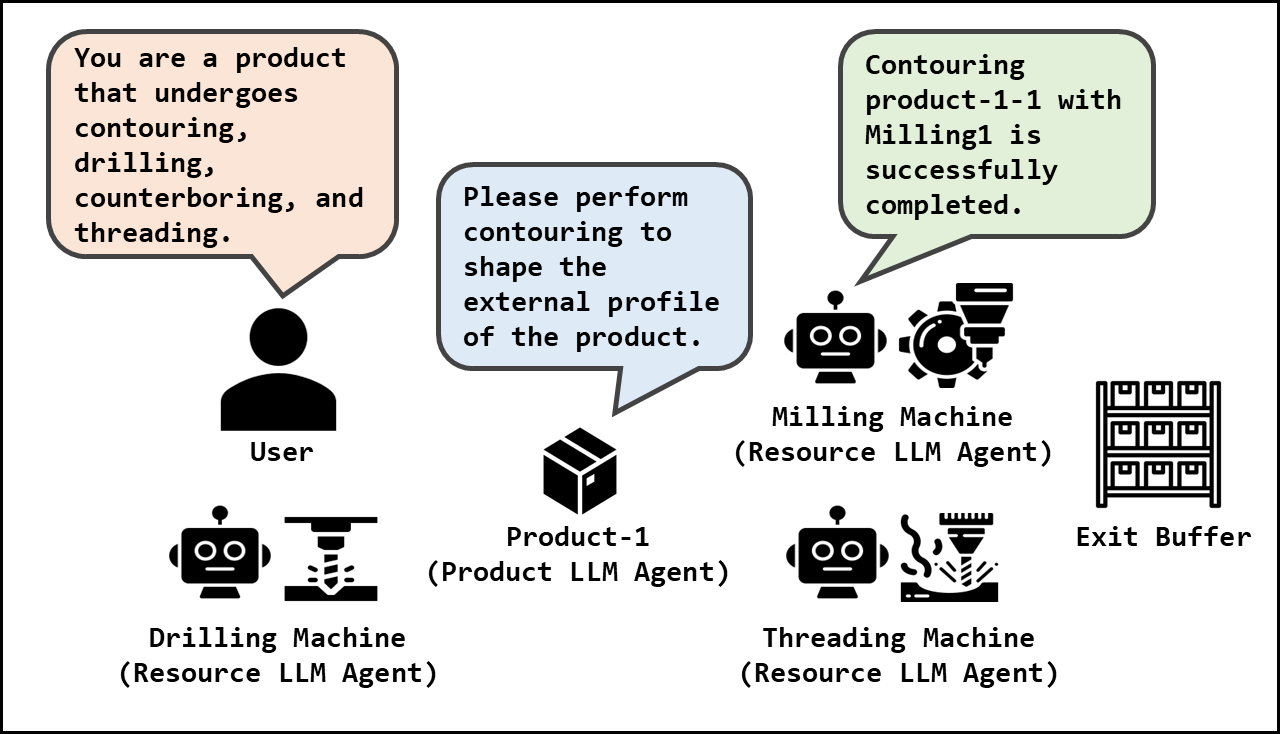}
    \caption{Case Study Setup}
    \label{fig:setup}
\end{figure}

\begin{figure}[t]
\smallskip
\smallskip
    \centering
    \includegraphics[width=.48\textwidth]{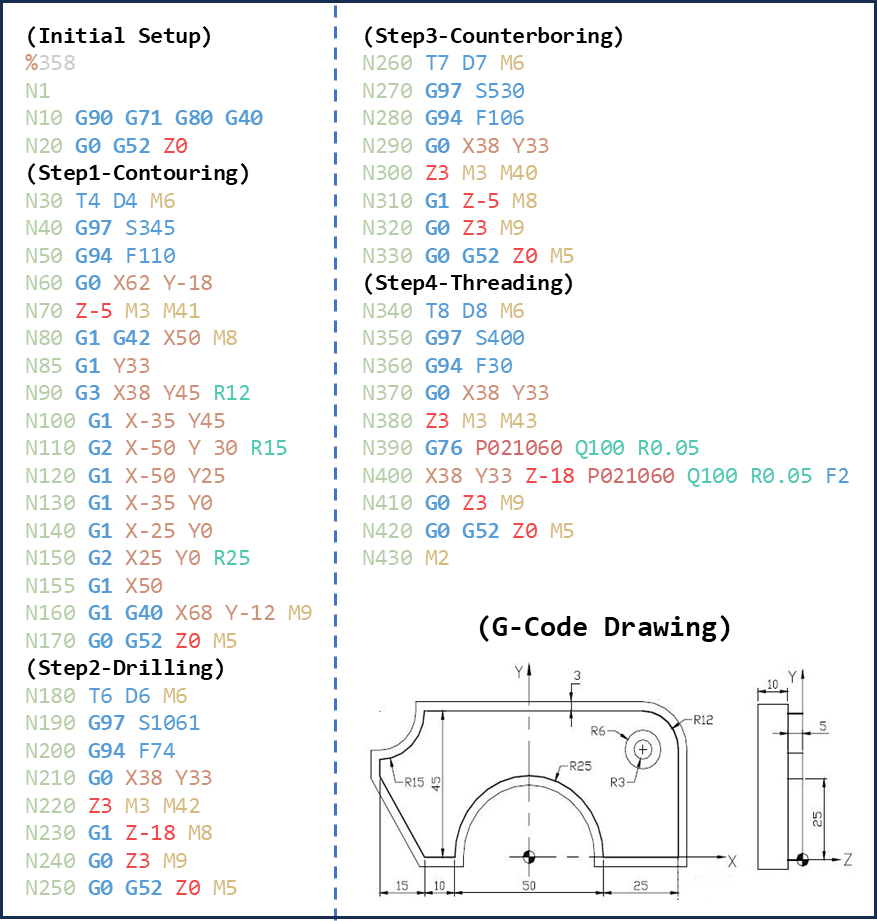}
    \caption{Case Study G-code Product Specification}
    \label{fig:specification}
\end{figure}

A simulated manufacturing environment includes three machines specialized for milling, drilling, and threading tasks, as highlighted in~\Cref{fig:setup}. To evaluate the system's adaptability, we structured the case study into two distinct processes: a 2-step process and a more comprehensive 4-step process. In the 2-step process, a prototype product, referred to as `product-1', is subjected to contouring and drilling operations. The 4-step process adds counterboring and threading operations, testing the system's capability to handle extended workflows. The G-code specifications are adapted from~\cite{HelmanCNC_2014}, as detailed in~\Cref{fig:specification}. Notably, explicit labels for G-code specifications (e.g. Step1-Contouring, Step2-Drilling) are not provided to the agent. Leveraging \gls{llm}'s capability to interpret and assign appropriate G-code based on tool specifications for each \gls{ra} demonstrates the system's potential to manage G-code assignments, demonstrating the capabilities of \glspl{llm}.

Each product has an \gls{llm}-enabled \gls{pa} which oversees high-level decision-making for its associated product. For instance, `product-1' is instructed for a 2-step process with the following instructions:

\begin{quote}
{\footnotesize
\texttt{
"Make sure to get contouring to shape the external profile of the product. Then, get a drilling operation to create precise holes. Proceed to Exit Buffer when completed. Your product name is 'product-1'. The G-code specifications are as follows: \%358 N1 N10 G90 G71 G80 G40..."
}
}
\end{quote}

Each LLM-enabled \gls{ra} handles decisions for its specific machine. `Milling1' is tasked with both contouring and counterboring to demonstrate how \gls{llm} can assign operations adaptively. `Drilling1' focuses on drilling, while `Threading1' manages threading. Machines are assigned tool numbers used in G-code to ensure \glspl{pa} provide relevant G-code to each \gls{ra}. For example, the `Milling1' machine receives this operational prompt:

\begin{quote}
{\footnotesize
\texttt{
"Perform milling operations as required, including contouring the external shape and counterboring for holes. Your machine name is Milling1. The tool number for contouring is T4. The tool number for counterboring is T6."
}
}
\end{quote}

This experiment investigates the \gls{llm}'s capability to identify and allocate correct manufacturing task and their corresponding G-code specifications. It aims to enhance the adaptability by showcasing natural language communication between \gls{pa} and 
\gls{ra}, simulating human-like interaction for task execution.

\subsection{Case Study Simulation and Results Analysis}
\label{subsec:case-study-results}

\begin{figure}
\smallskip
\smallskip
    \includegraphics[width=.48\textwidth]{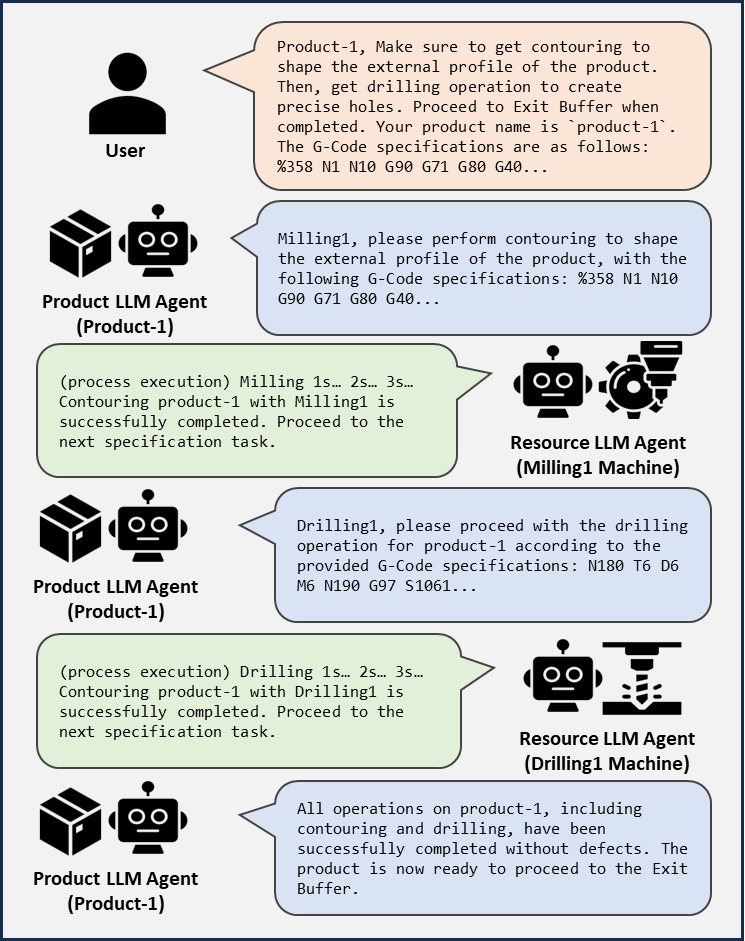}
    \caption{Agent Communication Result for 2-step Process}
    \label{fig:agentChat}
\end{figure}

Following the case study setup, simulations were conducted to evaluate the system, as shown in~\Cref{fig:agentChat}. In this example, the user initiates the workflow by instructing the `Product-1' agent and all \glspl{ra}, using ~\Cref{alg:agent_creation} from \Cref{subsec:initialization}. After initialization, the system follows the operational workflow in \Cref{subsec:workflow}. The `Product-1' agent first communicates the need for contouring to the 'Milling1' agent, providing only the contouring G-code section. After the `Milling1' agent completes the task, it informs the `Product-1' agent, indicating it is ready for the next operation. The process continues with the `Product-1' agent assigning the drilling operation to the `Drilling1' agent, which updates the `Product-1' agent after completion. When all operations are completed, the `Product-1' agent updates the user, indicating it is ready to move to the exit buffer. All agent communications are facilitated by the chat interaction protocol defined in~\Cref{alg:chat_interaction}, as outlined in~\Cref{subsec:capabilities}.

\subsubsection{Communication and Comprehension}
\label{subsec:communication}

Our case study evaluated the agents' ability to communicate and comprehend specific tasks. Using natural language processing, agents interacted in both 2-step and 4-step processes. In the 2-step process, the communication between the 'Product-1' agent and the `Milling1' and `Drilling1' agents showed the system's efficiency in handling straightforward task sequences and distributing G-code specifications. The 4-step process tested the agents' ability to manage complex sequences, with `Milling1' handling both contouring and counterboring. In this scenario, the `Product-1' agent coordinated contouring, drilling, counterboring, and threading, showing the agents' ability to interpret complex instructions.

\subsubsection{Accuracy and Performance}
\label{subsec:accuracy}

The data in~\Cref{table:success_rates} shows GPT-4 achieved 100\% success in 2-step and 86\% in 4-step processes across 50 trials each, with performance declining as task complexity increased. The 4-step process had a 14\% overall error rate. The detailed error breakdown in \Cref{table:error_analysis} shows that in the 4-step process, 43\% of errors were due to incorrect function calls and process failures, and 14\% were due to inaccurate G-code allocation. The increase in prompt length correlates with a rise in errors, highlighting the need for model training and improved prompt engineering. Addressing these issues will enhance task execution capabilities in \glspl{mas} for manufacturing, improving system adaptability and efficiency even in unexpected scenarios.

\subsubsection{Insights from Case Studies}
\label{subsec:insights}

The case study provided insights into the capabilities of \gls{llm}-enabled \glspl{mas} in manufacturing. The framework uses \gls{llm} to understand manufacturing tasks \textit{(Cn1)}, adapt to new product specifications \textit{(Cn2)}, and facilitate agent communication via natural language \textit{(Cn3)}. This enhances system flexibility and task management without relying on predetermined rules. However, accuracy issues with GPT-4 highlight the need for model improvement to manufacturing needs. Difficulties with incorrect function calls point to the necessity for better error management and validation. These results guide future research toward enhancing \gls{llm} integration, developing advanced agent communication protocols, and increasing system robustness.

\begin{table}[t]
\smallskip
\smallskip
\centering
\caption{Success Rates for Case Studies with GPT-4}
\label{table:success_rates}
\small
\begin{tabularx}{\columnwidth}{Xcc}
\hline
\textbf{Metric} & \textbf{2-Step} & \textbf{4-Step} \\
\hline
Success Rate                & 1      & .86      \\
Overall Error Rate          & 0       & .14       \\
- P1. Contouring Failure Rate     & 0       & 0       \\
- P2. Drilling Failure Rate       & 0       & .06       \\
- P3. Counterboring Failure Rate  & N/A     & .02       \\
- P4. Threading Failure Rate      & N/A     & .06       \\
\hline
\end{tabularx}
\end{table}

\begin{table}[t]
\centering
\caption{Detailed Error Analysis with GPT-4}
\label{table:error_analysis}
\small
\begin{tabularx}{\columnwidth}{Xcc}
\hline
\textbf{Error Type} & \textbf{2-Step} & \textbf{4-Step} \\
\hline
Incorrect Function Calls & 0 & .43 \\
Inaccurate G-code Allocation & 0 & .14 \\
Process Inexecution & 0 & .43 \\
\hline
\end{tabularx}
\end{table}

\section{Challenges and Limitations}
\label{sec:challenges}

\subsection{Performance}
\label{subsec:performance}
A significant challenge in manufacturing using \glspl{llm} is achieving the performance required for high-quality production. While effective in processing natural language, \glspl{llm} sometimes lack precision, as indicated in~\Cref{subsec:accuracy}. This issue raises concerns about the feasibility of relying on \glspl{llm} for tasks where errors can cause machine breakdown and low product quality. The proposed framework can enhance performance by applying fine-tuned \glspl{llm} on domain-specific data~\cite{xia2024leveraging}.

\subsection{Scalability}
\label{subsec:scalability}
As data volume and task complexity grow, \glspl{llm} may struggle to maintain accuracy. This limitation is particularly noticeable in situations where extensive historical data and detailed prompts are essential, as exemplified in the 4-step process in~\Cref{sec:casestudy}. The potential decrease in accuracy with larger datasets poses a challenge to scaling \gls{llm}-enabled multi-agent control across more complex manufacturing operations. A knowledge base for multi-agent manufacturing systems~\cite{lim2023ontology} should be integrated for practical industrial applications to provide \glspl{llm} with accurate runtime data and to enhance scalability.

\subsection{Reliability}
\label{subsec:reliability}
Reliability is key for integrating this method into real manufacturing scenarios. \glspl{llm} generating hallucinated responses or making false decisions raises safety concerns, as shown in~\Cref{sec:casestudy}. The \gls{llm} allocated inaccurate G-code or executed the incorrect function calls. Moreover, the reasoning provided by \glspl{llm} may require validation in practical manufacturing scenarios even if it seems logical. Therefore, it is essential to implement robust verification mechanisms, as well as incorporate real-time observability and monitoring capabilities. This approach ensures \glspl{llm}' outputs and reasoning are consistent and allow immediate adjustments with manufacturing standards and safety protocols. Joublin et al.~\cite{joublin2023copal} demonstrate a system for robotics that uses \glspl{llm} for task planning and replanning in response to errors, highlighting the importance of feedback loops for safety and reliability in complex manufacturing environments.

\section{Conclusion and Future Work}
\label{sec:conclusion}

In our research, we explored integrating \gls{llm} within \gls{mas} to enhance manufacturing processes.
Our research demonstrates that \gls{llm} improves task allocation efficiency and operational adaptability \textit{(Cn1)}, enables agents to adapt to evolving product specifications (e.g., G-code) \textit{(Cn2)}, and enhances agent communication and collaboration through advanced natural language processing \textit{(Cn3)}.
However, challenges like task execution performance, scalability, and reliability highlight the need for verification and monitoring. Additionally, the diversity in user instructions requires a strategy to ensure reliable task interpretation and execution.

Future work will focus on developing \gls{llm}-based communication protocols for agents to handle diverse manufacturing instructions and operator proficiency levels. This approach seeks to showcase the agents' capability to learn and adapt from historical data, enhancing their responsiveness to rapid industry changes. Additionally, efforts will be directed towards enhancing the accuracy and scalability of \gls{llm}. While this paper focused on common manufacturing processes (e.g. drilling, milling), future research will include less common processes with larger datasets to evaluate the framework's effectiveness. The goal is to ensure the efficient processing of diverse manufacturing tasks without impacting performance.


\section*{Acknowledgment}
We thank Felix Ocker for his valuable comments and suggestions.

\balance


\end{document}